\documentclass[11pt]{article}%
\usepackage{eurosym}
\usepackage{latexsym,delarray,multicol}
\usepackage{epsfig}
\usepackage{amsfonts}
\usepackage{amsthm}
\usepackage{amsmath}
\usepackage{amssymb}
\usepackage{revsymb}
\usepackage{graphicx}
\usepackage{blindtext}%
\newcommand{\nc}{\newcommand}
\nc{\nn}{\nonumber} \nc{\ep}{\varepsilon} \nc{\la}{\lambda}
\nc{\wht}{\widehat} \nc{\ov}{\overline} \nc{\ds}{\displaystyle}
\nc{\ts}{\textstyle}
\nc{\kro}{\left(}\nc{\kvo}{\left[}\nc{\fio}{\left\{}
\nc{\krz}{\right)}\nc{\kvz}{\right]}\nc{\fiz}{\right\}}
\newtheorem{theorem}{Theorem}

\newtheorem{corollary}[theorem]{Corollary}

\newtheorem{statement}{Statement}

\newtheorem{lemma}[theorem]{Lemma}

\begin{document}

\title{Codes approaching the Shannon limit with polynomial complexity per information
bit\smallskip}
\author{Ilya Dumer and Navid Gharavi\\University of California, Riverside, USA\\Email: dumer@ee.ucr.edu, navid.gharavi@email.ucr.edu}
\date{}
\maketitle

\begin{abstract}
We consider codes for channels with extreme noise that emerge in various
low-power applications. Simple LDPC-type codes with parity checks of weight 3
are first studied for any dimension $m\rightarrow\infty.$ These codes form
modulation schemes: they improve the original channel output for any $SNR>$
$-6$ dB (per information bit) and gain $3$ dB over uncoded modulation as $SNR$
grows. However, they also have a floor on the output bit error rate (BER)
irrespective of their length. Tight lower and upper bounds, which are
virtually identical to simulation results, are then obtained for BER at any
SNR. We also study a combined scheme that splits $m$ information bits into $b$
blocks and protects each with some polar code. Decoding moves back and forth
between polar and LDPC codes, every time using a polar code of a higher rate.
For a sufficiently large constant $b$ and $m\rightarrow\infty$, this design
yields a vanishing BER at any SNR that is arbitrarily close to the Shannon
limit of -1.59 dB. Unlike other existing designs, this scheme has polynomial
complexity of order $m\ln m$ per information bit.

\end{abstract}

\section{Introduction}


In this work, we address code design that protects information transmitted on
the AWGN channels with extreme noise. One particularly ubiquitous application
is the Internet of things (IoT). To efficiently employ it, prospective
standards \cite{stand-LTE} are supposed to achieve a $20$ dB reduction in
$snr$ per channel bit (below SNR denotes the signal-to-noise ratio per
information bit, and notation $snr$ implies channel outputs).

From the theoretical standpoint, we consider binary linear codes $C(n,k)$ of
length $n\rightarrow\infty$ and dimension $k$ used on the BI-AWGN channels
$\mathcal{N}(0,\sigma_{n}^{2})$ with noise power $\sigma_{n}^{2}%
\rightarrow\infty.$ To achieve a fixed signal-to-noise ratio $SNR=1/\left(
2\sigma_{n}^{2}R_{n}\right)  ,$ these codes must have the vanishing code rates
$R_{n}$ that have an order of $\sigma_{n}^{-2}.$ Moreover, the fundamental
Shannon limit shows that any such code may achieve the vanishing BERs only if
$SNR>\ln2$ (equivalently, this limit corresponds to $10\log_{10}\ln2=-1.5917$ dB).

The central problem here is to design a capacity-achieving sequence of codes
that have low decoding complexity and a rapidly declining BER. Currently, this
problem is far from solution. To date, most existing capacity-achieving codes
have code rates $R_{n}$ that decline exponentially in code dimension $m.$ In
turn, this yields an exponential growth in bandwidth and decoding complexity,
both proportional to $R_{n}^{-1}.$

For example, biorthogonal codes $C(2^{m-1},m)$ achieve the Shannon limit;
however, their code rate $R_{n}=m/2^{m-1}$ declines exponentially in $m.$ By
contrast, the output word error rates (WER) of these codes experience very
slow decline, which is only polynomial in blocklength $n$. In particular, for
the low $SNR\in(\ln2,4\ln2)$, codes $C(2^{m-1},m)$ have WER \cite{forney}
bounded from above by \vspace{0in}
\[
P_{m}=\exp\{-m(\sqrt{SNR}-\sqrt{\ln2})^{2}\}
\]
For a practically important range of $SNR\in\lbrack1,2]$ (which gives the
range of $[0,3]$ dB), long codes $C_{m}$ -- up to billions of bits -- still
have very high error rates $P_{m}.$ This is shown below for $m=18$ and $m=30$.
\smallskip%

\[%
\begin{tabular}
[c]{|l|l|l|l|}\hline
SNR (dBs) & 0 & 1 & 2\\\hline
$P_{18}$ & $.60$ & $.22$ & $.04$\\\hline
$P_{30}$ & $.43$ & $.08$ & $.0045$\\\hline
\end{tabular}
\ \
\]
Further analysis shows that concatenations of codes $C(2^{m-1},m)$ with the
outer RS codes or AG codes still have similar shortcomings, due to the fact
that codes $C(2^{m-1},m)$ should have length $n$ proportional to $\sigma
_{n}^{2}\rightarrow\infty.$ In summary, codes $C_{m}$ or their concatenations
fail to yield acceptable output error rates on the high-noise AWGN channels
with SNR of [0,2] dB for the blocks of length $n<10^{8}$.

As the second example, consider general RM codes or their bit-frozen subcodes.
Let $W_{m}$ be a sequence of the binary symmetric channels (BCH$_{p})$ with
transition error probabilities $p_{m}=(1-\epsilon_{m})/2$ such that
$\epsilon_{m}\rightarrow0$ as $m\rightarrow\infty.$ It is well known that
channels $W_{m}$ yield a sequence of vanishing capacities
\[
C_{m}\sim\epsilon_{m}^{2}/\ln4,\quad m\rightarrow\infty
\]
It was proven in \cite{abbe,shpilka-2016} that long low-rate RM codes
$RM(m,r)$ of order $r=o(m)$ and length $n=2^{m}$ approach the maximum possible
code rates $C_{m}$ on channels $W_{m}$ under the maximum-likelihood (ML)
decoding. Even in this case, code rates $R_{n}$ decline exponentially as
$m^{r}2^{-m}$ and require exponential decoding complexity.

Consider also the existing low-complexity algorithms known for RM codes
\cite{sid-1992}, \cite{loidr}, \cite{ming} or their bit-frozen subcodes
\cite{dum-2001,dum-2006}. For low $SNR<1$ dB, these algorithms yield high
error rates above $10^{-3}$ or require unacceptably large lists under
successive cancellation list (SCL) decoding.

Finally, consider polar codes \cite{ari} of rate $R_{n}\rightarrow0$ that
operate under growing noise power $\sigma_{n}^{2}\sim1/\left(  2SR_{n}\right)
$ for a fixed SNR $=S.$ One construction of such codes is considered in
\cite{dum-2017}. For $\sigma_{n}^{2}\rightarrow\infty,$ these codes begin with
a growing number $\mu\sim$ $\log_{2}\sigma_{n}^{2}$ of upgrading channels and
employ long repetition codes $B(2^{\mu},1)$ or RM codes $C(2^{\mu},m+1)$. This
design again results in a rapid complexity increase as $\sigma_{n}%
^{2}\rightarrow\infty$. To advance polar design, it is important to analyze
how polar codes of length $n\rightarrow\infty$ operate within a vanishing
margin $\varepsilon_{n}\rightarrow0$ to the Shannon limit. One particular
problem is to derive the trade off between the BER and code complexity arising
when $\varepsilon_{n}\rightarrow0$.

For moderate lengths, one efficient construction of \cite{hessam} concatenates
repetition code of length 4 with a (2048,40) polar code. The resulting code
has WER of $.002$ at the SNR of 2 dB and improves the NB-IoT standard
\cite{stand-LTE} by 1 dB. Another recent design \cite{gate1} yields WER of
$.0007$ for the similar parameters. Below we improve asymptotic performance of
codes \cite{gate1} with a new design and analytical tools. Our main statement
is as follows.

\begin{statement}
\label{th:1}There exist codes $\widehat{C}_{m}$ of dimension $k\rightarrow
\infty$ and length $O(k^{2})$ that have complexity of order $\mathcal{O(}%
k^{2}\log k)$ and limit $BER$ to the order of $\exp\{-c_{SNR}\sqrt{k}\},$
where $c_{SNR}>0$ depends on SNR and is positive for any $SNR$ above the
Shannon limit of $\ln2.$
\end{statement}

Statement \ref{th:1} is predicated on our "weak-independence" assumption
discussed in Section \ref{s:prob}.\vspace{-0.01in}\vspace{-0.02in}

\section{Basic construction and its decoding algorithm}

Our basic code - which we denote $C_{m}$ - has generator matrix $G_{m}%
=[I_{m}|J_{m}]$, where $I_{m}$ is an $m\times m$ identity matrix and $J_{m}$
is an $m\times\left(  _{\,2}^{m}\right)  $ matrix that includes all columns of
weight 2. Clearly, $n=\left(  _{\,\,\,2}^{m+1}\right)  $ and $k=m{.}$ Let
$a_{(s)}$ be any codeword generated by $s$ rows of $G_{m}.$ Note that every
row in $J_{m}$ has weight $m-1,$ every two rows have a single common $1$, and
every $s\geq2$ rows have $\left(  _{2}^{s}\right)  $ common $1$s. Any codeword
$a_{(s)}$ that has weight $s$ in $I_{m}$ has overall weight
\begin{equation}
w_{s}=ms-2\left(  _{2}^{s}\right)  =s(m-s+1) \label{spectra}%
\end{equation}
Thus, code $C_{m}$ has distance $m,$ which is achieved if $s=1,m$.

Let $[i,j]=[j,i]$ denote code positions in $G_{m},$ where $0\leq i\neq j\leq
m.$ Encoder $aG_{m}$ receives a string $a=\left(  a_{0,1},...,a_{0,m}\right)
$ of $m$ information bits and adds $\left(  _{\,2}^{m}\right)  $ parity bits
$a_{1,2},...,a_{m-1,m}$ such that $a_{i,j}=a_{0,i}+a_{0,j}.$ Note that
encoding has complexity $\mathcal{O(}n).$

Let code $C_{m}$ of rate $R=2/(m+1)$ be used on an AWGN channel with p.d.f.
$\mathbb{N}(0,\sigma^{2})$ and constant $SNR=\left(  2\sigma^{2}R\right)
^{-1}$ per information bit. In the sequel, it will be convenient for us to use
a constant $c=4(SNR)$. We use a map $\{0,1\}\rightarrow\{\pm1\}$ for each
transmitted symbol $a_{i,j}$, where $0\leq i\neq j\leq m.$ Then the parity
checks $a_{i,j}$ form the real-valued products\vspace{0in}
\begin{equation}
a_{0,i}=a_{0,j}a_{i,j}\vspace{0in} \label{4}%
\end{equation}
Let an all-one codeword $1^{n}$ be transmitted. Then the received symbols
$y_{i,j}\equiv y_{j,i}$ form independent Gaussian R.V. $\mathbb{N}%
(1,\sigma^{2}).$ We will use rescaled r.v. $z_{i,j}=\delta y_{i,j,}$where
$\delta=1/\left(  \sigma^{2}+1\right)  =c/\left(  m+c+1\right)  .$ It is easy
to verify that this scaling gives power moments $x_{0}=E(z_{i,j})$ and
$\sigma_{0}^{2}=E(z_{i,j}^{2})$ such that\vspace{0in}
\begin{equation}
x_{0}=\sigma_{0}^{2}=\delta\label{eq-0}%
\end{equation}
Given some $z_{i,j},$ an input $a_{i,j}=1$ has posterior probability\vspace
{-0.02in}
\[
q_{i,j}\triangleq\Pr\{1\mid z_{i,j}\}=1/(\exp\left(  -2z_{i,j}\right)  +1).
\]
Decoding algorithm $\Psi_{soft}(z)$ described below employs two closely
related quantities, the log-likelihoods (l.l.h.) $h_{i,j}$\ and the
\textquotedblleft probability offsets" $u_{i,j}:$\vspace{0in}%
\begin{align}
h_{i,j}  &  =\ln[q_{i,j}]-\ln[1-q_{i,j}]=2z_{i,j}\label{h}\\
u_{i,j}  &  =2q_{i,j}-1=\tanh(z_{i,j}) \label{g}%
\end{align}
Given the offsets $u_{0,j}$ and $u_{i,j}$ in (\ref{4}), it is easy to verify
that symbol $a_{0,i}$ has offset $u_{0,i}=u_{0,j}u_{i,j}.$ Also,
$u_{i,j}=\tanh(z_{i,j})=\tanh(h_{i,j}/2)$. Function $\tanh(x)$ has derivatives
$\tanh^{\prime}(0)=1$ and $\tanh^{\prime\prime}(0)=0$ at $x=0.$ Thus, for the
vanishing values of $z_{i,j}\rightarrow0,$ \vspace{0in}
\begin{equation}
u_{i,j}=z_{i,j}+o(z_{i,j}^{2})=h_{i,j}/2+o(h_{i,j}^{2}) \label{gh}%
\end{equation}
Algorithm $\Psi_{soft}$ performs several steps of belief propagation. Unlike
conventional algorithms, we estimate only information bits $a_{0,i}$. We will
show that $\Psi_{soft}$ requires $L\sim\ln m\,/\ln c$ iterations to achieve
the best performance$.$

For every step $\ell=1,...,L$ and every symbol $a_{0,i},$ consider its $j$-th
parity check $a_{0,i}=a_{0,j}a_{i,j}$ of (\ref{4})$.$ To re-evaluate $a_{0,i}%
$, we introduce the offset $u_{i\,|\,\ell}(j)$ of the symbol $a_{0,j}$ used in
this parity check. Then the estimate $u_{i,j}u_{j\,|\,\ell}(j)$ re-evaluates
symbol $a_{0,i}$ via the product $a_{0,j}a_{i,j}$. We then obtain the l.l.h.
$h_{i\,|\,\ell+1}(j)$ of the $j$-th parity check using transforms (\ref{h})
and (\ref{g}). Next, the sum of l.l.h. $h_{i\,|\,\ell+1}(j)$ gives the
compound estimate $h_{i\,|\,\ell+1}$ of the symbol $a_{0,i}$. Finally, we
derive the partial l.l.h. $h_{j\,|\,\ell+1}(i)$ of the symbol $a_{0,i}$ that
will be used in the next round to estimate $a_{0,j}$ via its $i$-th parity
check $a_{0,j}=a_{0,i}a_{i,j}$. This excludes the intrinsic information
$h_{i\,|\,\ell+1}(j)$ that symbol $a_{0,j}$ already used in round $\ell$. Our
recalculations begin with the original estimates $u_{i\,|\,0}(j)\triangleq
u_{0,i}.$ Round $\ell$ of $\Psi_{soft}$ is done as follows.%

\begin{equation}
\frame{$%
\begin{array}
[c]{l}%
\text{For all }i,j\in\{1,...,m\}\text{ and }j\neq i:\smallskip\\
A.\text{ Derive quantities }u_{i\,|\,\ell+1}(j)=u_{i,j}u_{i\,|\,\ell}(j)\text{
}\vspace{0.02in}\vspace{0.02in}\\
\quad\text{and }h_{i\,|\,\ell+1}(j)=2\tanh^{-1}\left[  u_{i\,|\,\ell
+1}(j)\right]  .\vspace{0.02in}\vspace{0.02in}\\
B.\text{ Derive quantities }h_{i\,|\,\ell+1}=\sum\nolimits_{j}h_{i\,|\,\ell
+1}(j)\vspace{0.02in}\vspace{0.02in}\\
\quad\text{and }h_{j\,|\,\ell+1}(i)=h_{i\,|\,\ell+1}-h_{i\,|\,\ell
+1}(j)\vspace{0.02in}\vspace{0.02in}\\
C.\text{ If }\ell<L,\text{ find }u_{i\,|\,\ell+1}(j)=\tanh(h_{i\,|\,\ell
+1}(j)/2).\vspace{0.02in}\vspace{0.02in}\text{ }\\
\quad\text{Go to A with }\ell:=\ell+1.\text{ If }\ell=L:\vspace{0.02in}\text{
}\vspace{0.02in}\\
\quad\text{estimate BER }\tau_{_{L}}=\frac{1}{m}\sum_{i}\Pr\{h_{i\,|\,L}%
<0\};\vspace{0.02in}\vspace{0.02in}\text{ }\\
\quad\text{output numbers }h_{i\,|\,L}\text{ and }a_{i}=\text{sign
}(h_{i\,|\,L}).
\end{array}
$ } \label{soft}%
\end{equation}
To estimate the complexity of $\Psi_{soft}$, note that Step A uses at most $n$
multiplications and $n$ two-way conversions $u\leftrightarrow h.$ Step $B$
calculates the sums $h_{i\,|\,\ell+1}$ using $m$ operations for each $i$. It
also requires $2n$ operations to derive the residual sums $h_{i\,|\,\ell
+1}(j)$ and their offsets $u_{i\,|\,\ell+1}(j)$ for all pairs $i,j$. Then the
overall complexity has the order $\mathcal{O(}n)$ for every iteration $\ell.$
Assuming that we have $L=\mathcal{O(}\log m)$ iterations, we obtain complexity
$\mathcal{O(}n\log n).$

\section{Lower bounds for BER of codes $C_{m}$}

We will now study the output BER of codes $C_{m}.$ We first show that long
codes $C_{m}$ fail to achieve BER $P_{c}\rightarrow0$ for any $SNR=c/4$ even
if they employ ML decoding$.$ This is similar to the uncoded modulation (UM).
Let
\[
Q(x)=(2\pi)^{-1/2}\int\nolimits_{x}^{\infty}\exp\{-y^{2}/2)dy
\]
Assume that an all-one codeword $1^{n}$ (formerly, a $0^{n}$ codeword in
$\mathbb{F}_{2}^{n})$ is transmitted and $z=(z_{i,j})$ is received. Consider
the sets of positions $I_{0}=(0,j\,|\,j\neq0,1)$ and $I_{1}=(0,j\,|\,j\neq
0,1).$ For any vector $z,$ we will define the corresponding r.v.
\[
Y_{0}=\sum\nolimits_{j\neq0,1}z_{0,j},\;\;Y_{1}=\sum\nolimits_{j\neq
0,1}z_{1,j}%
\]
Below we use asymptotic pdfs as $m\rightarrow\infty.$ Then r.v. $z_{i,j}$ have
asymptotic pdf $\mathbb{N}(\delta,\delta)$. It is also easy to verify that
r.v. $Z_{i}=\sum\nolimits_{j}z_{i,j},$ $Y_{0},$ and $Y_{1}$ have asymptotic
pdf $\mathbb{N}\mathcal{(}c,c).$\vspace{0.02in}

Codewords of minimum weight in $C_{m}$ include $m$ generator rows $g^{(p)},$
$p=1,...,m,$ of the generator matrix $G_{m}$ and their sum $g^{(0)}%
=g^{(1)}+...+g^{(m)}.$ Under ML decoding, any two-word code $\{1^{n},$
$g^{(p)}\},$ has BER
\begin{equation}
P_{c}=\Pr\left\{  Y_{1}<0\right\}  =Q\left(  \frac{m\delta-\delta}%
{\sqrt{m(\delta-\delta^{2})}}\right)  \sim Q\left(  \sqrt{c}\right)
\label{event-1}%
\end{equation}
Here we write $f(m)\sim g(m)$ if $\lim f(m)/g(m)=1$ as $m\rightarrow\infty.$
Similarly, we use notation $f(m)\gtrsim g(m)$ if $\lim f(m)/g(m)\geq1.$

\begin{theorem}
\label{lm:0} Let codes $C_{m}$ be used on an AWGN channel with an SNR of $c/4$
per information bit. Then for $m\rightarrow\infty,$ ML decoding of codes
$C_{m}$ has BER\vspace{0.0in}
\begin{equation}
p_{ML}(c)\gtrsim2P_{c}(1-P_{c})=2Q\left(  \sqrt{c}\right)  -2Q^{2}\left(
\sqrt{c}\right)  \label{event-2}%
\end{equation}

\end{theorem}

\noindent\textit{Proof. } Without loss of generality, we consider BER of
symbol $a_{0,1}$. In essence, we prove that ML decoding gives $a_{0,1}=-1$ if
so does one of the codes $\{1^{n},$ $g^{(p)}\}$ for $p=0,1.$ All received
vectors $z$ form four disjoint subsets $U=A,B,C,D,$ where\vspace{0in}
\begin{align}
A  &  =\{z\,|\,Y_{0}<0,\;Y_{1}>0\},\;B=\{z\,|\,Y_{0}>0,\;Y_{1}<0\}\label{A}\\
C  &  =\{z\,|\,Y_{0}>0,\;Y_{1}>0\},\;D=\{z\,|\,Y_{0}<0,\;Y_{1}<0\} \label{D}%
\end{align}
Clearly, $\Pr\{A\}=\Pr\{B\}=P_{c}(1-P_{c}).$ We will prove that $p_{ML}%
(c)\gtrsim\Pr\{A\}+\Pr\{B\}.$

Two vectors $g^{(p)},$ $p=0,1,$ have supports $J_{p}=\{(p,j)\},$ where
$j\in\{0,...,m\}\backslash\left\{  p\right\}  .$ For any $z$, consider bitwise
products $g^{(p)}z$ that flip symbols of $z$ on the supports $J_{p}.$
Then\vspace{0.0in}%

\begin{equation}
g^{(0)}A=C,\;g^{(1)}A=D,\;g^{(0)}B=D,\;g^{(1)}B=C \label{event-3}%
\end{equation}
Let $z$ be decoded into some $a(z)\in$ $C_{m}$ and let $a_{0,1}(z)$ be the
first symbol of $a(z)$. We decompose each set $U$ into\vspace{0in}
\[
U_{+}=\{z\in U:a_{0,1}(z)=1\},\;U_{-}=\{z\in U:a_{0,1}(z)=-1\}
\]
Note that $a\left(  g^{(p)}z\right)  =g^{(p)}a(z).$ Then
\begin{align}
g^{(0)}A_{+}  &  =C_{-},\;\;g^{(1)}A_{+}=D_{-}\label{event-4}\\
g^{(1)}B_{+}  &  =C_{-},\;\;g^{(0)}B_{+}=D_{-}\nonumber
\end{align}
Conditions (\ref{event-3}) and (\ref{event-3}) show that maps $g^{(0)}$ and
$g^{(1)}$ flip full sets $U$ and there subsets $U_{+}$ and $U_{-}.$

In the next step, we remove the first symbol $a_{0,1}$ from each vector $z$
and obtain four sets $U^{\prime}=A^{\prime},B^{\prime},C^{\prime},D^{\prime}$
with a punctured symbol $a_{0,1}.$ Let $U_{+}^{\prime}$ and $U_{-}^{\prime}$
denote the punctured subsets of $U_{+}$ and $U_{-}.$ Below we show in Lemma
\textit{\ref{lm:1-1} } that the maps $g^{(0)}$ and $g^{(1)}$ cannot reduce the
probability of the sets $A^{\prime}+B^{\prime}.$ Namely,
\begin{align}
\Pr\left\{  C_{-}^{\prime}\right\}  +\Pr\left\{  D_{-}^{\prime}\right\}   &
\geq2\Pr\left\{  A_{+}^{\prime}\right\} \label{flip-4}\\
\Pr\left\{  C_{-}^{\prime}\right\}  +\Pr\left\{  D_{-}^{\prime}\right\}   &
\geq2\Pr\left\{  B_{+}^{\prime}\right\}  \label{flip-3}%
\end{align}
Finally, consider $p_{ML}(c)\equiv\sum\nolimits_{U}\Pr\left\{  U_{-}\right\}
.$ We then prove in Lemma \textit{\ref{lm:1-a} } that removing one bit
$a_{0,1}$ has immaterial impact on $\Pr\{U\}$ as $m\rightarrow\infty,$ so that
$\Pr\{U\}\sim\Pr\{U^{\prime}\}.$ Then
\[
p_{ML}(c)=\sum\nolimits_{U}\Pr\left\{  U_{-}\right\}  \sim\sum\nolimits_{U}%
\Pr\left\{  U_{-}^{\prime}\right\}
\]
We can now use (\ref{flip-4}) and (\ref{flip-3}), which gives \vspace{0.0in}%
\begin{align*}
p_{ML}(c)  &  \sim\Pr\left\{  A_{-}^{\prime}\right\}  +\Pr\left\{
B_{-}^{\prime}\right\}  +\Pr\left\{  C_{-}^{\prime}\right\}  +\Pr\left\{
D_{-}^{\prime}\right\} \\
&  \geq\Pr\left\{  A_{-}^{\prime}\right\}  +\Pr\left\{  B_{-}^{\prime
}\right\}  +\Pr\left\{  A_{+}^{\prime}\right\}  +\Pr\left\{  B_{+}^{\prime
}\right\} \\
&  =\Pr\left\{  A^{\prime}\right\}  +\Pr\left\{  B^{\prime}\right\}
\end{align*}
Thus, we obtain (\ref{event-2}). \textit{ }\hfill
${\dimen0=1.5ex\advance\dimen0by-0.8pt\relax\blacksquare}$\vspace{0.05in}

\begin{lemma}
\label{lm:1-1} Punctured sets $U^{\prime}=A^{\prime},B^{\prime},C^{\prime
},D^{\prime}$ satisfy inequalities (\ref{flip-4}) and (\ref{flip-3}%
).\vspace{0.02in}
\end{lemma}

\noindent\textit{Proof. }Recall that $1^{n}$ is the transmitted vector. In
this case, the set $C$ has the highest probability among all sets $U$, while
$D$ is the least likely. We now can establish stronger conditions. In essence,
we show that the transition $A\mapsto C$ (or $B\mapsto C)$ produces a greater
increase $\Pr(C)-\Pr(A)$ than the drop $\Pr(A)-\Pr(D)$ required in transition
$A\mapsto D.$

We say that any $x\in A^{\prime},B^{\prime}$ is a $\left(  \theta,\rho\right)
$ vector if $Y_{0}=\theta,\;\;Y_{1}=\rho.$ According to (\ref{A}), any $x\in
A$ has $\theta<0,$ $\rho>0$, whereas it is vice versa for $x\in B.$

Recall that r.v. $Y_{0},$ $Y_{1}$ have asymptotic pdf $\mathbb{N}%
\mathcal{(}c,c).$ (The exact pdf is $\mathbb{N}\mathcal{(}c\lambda
,c\lambda-c\delta\lambda)).$ Consider $\left(  \theta,\rho\right)  $-vectors
$x\in A.$ On the subset $I_{0}=\left\{  (0,j)\right\}  ,$ these vectors $x$
have pdf
\[
p\left(  \theta\right)  \sim\left(  2\pi c\right)  ^{-1/2}e^{-\left(
\theta-c\right)  ^{2}/2c}%
\]
For any $x,$ the transform $g^{(0)}x$ only flips symbols $x_{0,j}$ thus
replacing p.d.f. $p\left(  \theta\right)  $ on the set $I_{0}$ with $p\left(
-\theta\right)  .$ This gives the ratio
\[
r\left(  \theta\right)  =p\left(  -\theta\right)  /p\left(  \theta\right)
=e^{-2\theta}%
\]
The other transform $g^{(1)}x$ of any $\left(  \theta,\rho\right)  $-vector
$x$ flips symbols $x_{1,j}.$ Then we obtain the ratio
\[
r\left(  \rho\right)  =p\left(  \rho\right)  /p\left(  -\rho\right)
=e^{-2\rho}%
\]
Now we consider two vectors from $A_{+},$ namely, $x=x\left(  \theta
,\rho\right)  $ and $y=y\left(  -\rho,-\theta\right)  $. Then $g^{(0)}x\in C$
and $g^{(1)}x\in D.$ The same inclusion holds for vector $y.$ Also, both
vectors $x$ and $y$ have the same pdf $p(x)=p(y)=p$ generated on the sets
$I_{0}$ and $I_{1},$ since both r.v. $Y_{0}$ and $Y_{1}$ have the same
distribution. We can now estimate the total pdf of vectors $g^{(p)}x$ and
$g^{(p)}y$ as follows
\begin{align*}
p\left(  g^{(0)}x\right)  +p\left(  g^{(1)}x\right)   &  =\left(  e^{-2\theta
}+e^{-2\rho}\right)  p\\
p\left(  g^{(0)}y\right)  +p\left(  g^{(1)}y\right)   &  =\left(  e^{2\theta
}+e^{2\rho}\right)  p
\end{align*}
Since $\exp\{-2a\}+\exp\{2a\}\geq2$ for any $a,$ we can reduce the latter
equalities to%
\[
2\sum\nolimits_{p=1,2}p\left(  g^{(p)}x\right)  +p\left(  g^{(p)}y\right)
\geq4p
\]
This immediately leads to inequality (\ref{flip-4}). Inequality (\ref{flip-3})
is identical if we replace $A_{+}$ with $B_{+}.$ Other inequalities of the
same kind can be obtained if we consider subsets $A^{\prime},B^{\prime}$ (or
$A_{-}^{\prime},B_{-}^{\prime}).$ \textit{ } \textit{ }\hfill
${\dimen0=1.5ex\advance\dimen0by-0.8pt\relax\blacksquare}$

{We} now prove that removing position $\left(  0,1\right)  $ is immaterial for
our proof.

\begin{lemma}
\label{lm:1-a} Any set $U$ and its one-bit puncturing $U^{\prime}$ satisfy
asymptotic equality $\Pr\{U\}\sim\Pr\{U^{\prime}\}.$
\end{lemma}

\noindent\textit{Proof. }Note that r.v. $z_{0,1}$ has pdf $\mathbb{N}%
(\delta,\delta),$ where $\delta\sim c/m\rightarrow0$ as $m\rightarrow\infty,$
whereas r.v. $Y_{0}$ (or $Y_{1})$ has pdf $\mathbb{N}(\delta,\delta).$ Let
$r=\sqrt{c/m}\ln m$ and $r^{\prime}=r\ln m.$ Then with probability tending to
1, we have the following conditions:%
\begin{equation}
z_{0,1}\in\lbrack-r,r],\;Y_{0}\notin\lbrack-r^{\prime},r^{\prime}]
\label{z-prob}%
\end{equation}
Thus, $\Pr\{z_{0,1}/Y_{0}\rightarrow0\}\rightarrow1$ as $m\rightarrow\infty.$
Now we see that equalities $\Pr\{U\}\sim\Pr\{U^{\prime}\}$ hold for any set
$U$ or $U_{+}$ or $U_{-}$ as $m\rightarrow\infty.$\textit{ } \textit{ }%
\hfill${\dimen0=1.5ex\advance\dimen0by-0.8pt\relax\blacksquare}$%
\vspace{-0.02in}

\section{Probabilistic Bounds for BP decoding\label{s:prob}\textit{ }}

Our next goal is to study BP algorithm $\Psi_{soft}$ of (\ref{soft}). We first
slightly expand on our notation. We say that events $U_{m}$ hold with high
probability $P_{m}$ if $P_{m}\rightarrow1$ as $m\rightarrow\infty.$ Let
$\mathbb{N}(a,b)$ denote the pdf of a Gaussian r.v. that has mean $a$,
variance $b,$ and the second power moment $a^{2}+b$. Consider a sequence of
Gaussian r.v. $x_{m}$ that have pdf $\mathbb{N}(a,b_{m}),$ where
$b_{m}=b(1+\theta_{m}),$ $b>0$ is a constant, and $\theta_{m}\rightarrow0$ as
$m\rightarrow\infty.$ Consider also any sequence $t_{m}$ such that
$t_{m}=o(\theta_{m}^{-1/2}).$ Then $\Pr\{x_{m}>t_{m}\}\sim Q(\left(
t_{m}-a\right)  b^{-1/2})$ and we write $\mathbb{N}(a,b_{m})\sim$
$\mathbb{N}(a,b)$.

Consider also r.v. $z_{i,j}$ that has pdf asymptotic $\mathbb{N}(\delta
,\delta)$ as $m\rightarrow\infty.$ Then restriction (\ref{z-prob}) shows that
with high probability $z_{i,j}\rightarrow0.$ Then equality (\ref{gh}) shows
that $u_{i,j}=z_{i,j}+o(z_{i,j}^{2})\sim z_{i,j}.$ Thus, we will replace r.v.
$u_{i,j}$ in algorithm $\Psi_{soft}$ with $z_{i,j}.$

To derive analytical bounds, we will slightly simplify algorithm $\Psi_{soft}$
and assign the same value $h_{i\,|\,\ell+1}(j)=h_{i\,|\,\ell+1}$ for all $j$
instead of different assignments $h_{i\,|\,\ell+1}(j):=h_{i\,|\,\ell
+1}-h_{j\,|\,\ell+1}(i).$ It can be shown that this change is immaterial for
our asymptotic analysis. It also makes very negligible changes even on the
short blocks $C.$ The simplified version of the algorithm $\Psi_{soft}$ -
described below - begins with the initial assignment $u_{j\,|\,0}=z_{0,j}$ in
round $\ell=0.$ We will perform $L=2\ln m/\ln c$ rounds. In round $\ell,$
$\Psi_{soft}$ proceeds as follows.%

\begin{equation}
\frame{ $%
\begin{array}
[c]{l}%
A.\text{ Derive quantities }u_{i\,|\,\ell+1}(j)=z_{i,j}u_{j\,|\,\ell}\text{
}\vspace{0.02in}\vspace{0.02in}\\
\quad\text{and }h_{i\,|\,\ell+1}(j)=2\tanh^{-1}\left[  u_{i\,|\,\ell
+1}(j)\right]  .\vspace{0.02in}\vspace{0.02in}\\
B.\text{ Derive quantities }h_{i\,|\,\ell+1}=\sum\nolimits_{j}h_{i\,|\,\ell
+1}(j)\vspace{0.02in}\vspace{0.02in}\\
C.\text{ If }\ell<L,\text{ find }u_{i\,|\,\ell+1}=\tanh(h_{i\,|\,\ell
+1}/2).\vspace{0.02in}\vspace{0.02in}\text{ }\\
\quad\text{Go to A with }\ell:=\ell+1.\text{ If }\ell=L:\vspace{0.02in}\text{
}\vspace{0.02in}\\
\quad\text{estimate BER }\tau_{_{L}}=\frac{1}{m}\sum_{i}\Pr\{h_{i\,|\,L}%
<0\};\vspace{0.02in}\vspace{0.02in}\text{ }\\
\quad\text{output numbers }h_{i\,|\,L}\text{ and }a_{0,i}=\text{sign
}(h_{i\,|\,L}).
\end{array}
$} \label{soft-1}%
\end{equation}
To derive analytical bounds, we will also assume that different r.v.
$h_{i\,|\,\ell}$ are \textquotedblleft weakly dependent". Namely, we call r.v.
$\xi_{1},...,\xi_{m}$ weakly dependent if for $m\rightarrow\infty,$ we have
asymptotic equality
\[
E(\xi_{i}\,|\,\xi_{j_{1}},...,\xi_{j_{b}})\rightarrow E(\xi_{i})
\]
for any constant $b,$ index $i,$ and any subset $J=\{j_{1},...,j_{b}\}$ such
that $i\notin J$. In particular, we will assume that the conditional moment
$E(h_{i\,|\,\ell+1}\,|\,h_{j_{1}\,|\,\ell},...,h_{j_{b}\,|\,\ell})$ tends to
the unconditional moment $E(h_{i\,|\,\ell+1}).$ This assumption does not
necessarily hold if $b$ is a growing number$.$ However, in our case, r.v.
$h_{i\,|\,\ell+1}$ includes $m-1$ different summands $h_{i\,|\,\ell+1}(j)$ for
all $j\neq i.$ On the other hand, only one related term $h_{j\,|\,\ell}(i)$ is
included in each sum $h_{j\,|\,\ell}$ for any $j\in J.$ (Both terms include
the same factor $u_{i,j}$ used to evaluate symbols $a_{0,i}$ and $a_{0,j}$ in
parity check (\ref{4})). The above assumption is also corroborated by the
simulation results, which essentially coincide with the theoretical bounds
derived below (see Fig. \ref{fig:BER}, in particular).

Our goal is to derive BER $P_{soft}(c)=\lim\tau_{_{L}}$ for $\Psi_{soft}$ as
$L,m\rightarrow\infty$. Given $c>0$, consider the equation
\begin{equation}
x=\frac{1}{\sqrt{2\pi}}\int_{-\infty}^{\infty}\tanh(t\sqrt{xc})e^{-\left(
t-\sqrt{xc}\right)  ^{2}/2}dt \label{eq-1}%
\end{equation}
In Lemma \ref{lm:1-4}, we will show that for $c\leq1$ equation (\ref{eq-1})
has a single root $x=0.$ For $c>1,$ (\ref{eq-1}) has the root $x=0$ and two
other roots $x_{\ast}$ and $-x_{\ast},$ where $x_{\ast}\in(0,1).$

For any $\ell=0,1,...,$ $L$ and any $m\rightarrow\infty,$ we introduce
parameter $c_{\ell}=c^{\left(  \ell+1\right)  /2}.$ We then derive
probabilities $P_{\ell}$ using recursion $P_{\ell+1}=S_{\ell}+P_{\ell}T_{\ell
}$, where
\begin{align}
S_{\ell}  &  =(2\pi)^{-1/2}\int_{-\infty}^{\infty}Q(c_{\ell}t)e^{-(t-c_{\ell
})^{2}/2}dt\label{q-ell}\\
T_{\ell}  &  =(2\pi)^{-1/2}\int_{-\infty}^{\infty}Q(c_{\ell}t)\left(
e^{-(t+c_{\ell})^{2}/2}-e^{-(t-c_{\ell})^{2}/2}\right)  dt \label{p-0}%
\end{align}
and $P_{0}=Q(\sqrt{c}).$ For any $\ell,$ probabilities $P_{\ell}$ depend on
$c$ only. We will also show that quantities $P_{\ell}$ converge exponentially
fast as $\ell\rightarrow\infty.$ Let $P_{\infty}=\lim\nolimits_{\ell
\rightarrow\infty}P_{\ell}.$ We can now establish the asymptotic value of BER
as $m\rightarrow\infty$.

\begin{theorem}
\label{lm:4} Let codes $C_{m}$ be used on an AWGN channel with an SNR $c/4$
per information bit. For $m\rightarrow\infty$ and $c\leq1,$ algorithm
$\Psi_{soft}$ has BER $P_{soft}(c)\rightarrow1/2.$\ For $c>1,$
\begin{equation}
P_{soft}(c)\sim\left(  1-P_{\infty}\right)  Q\left(  \sqrt{x_{\ast}c}\right)
+P_{\infty}(1-Q\left(  \sqrt{x_{\ast}c}\right)  ) \label{p-soft}%
\end{equation}

\end{theorem}

\begin{figure}[ptbh]
\centering\includegraphics[scale=0.4]{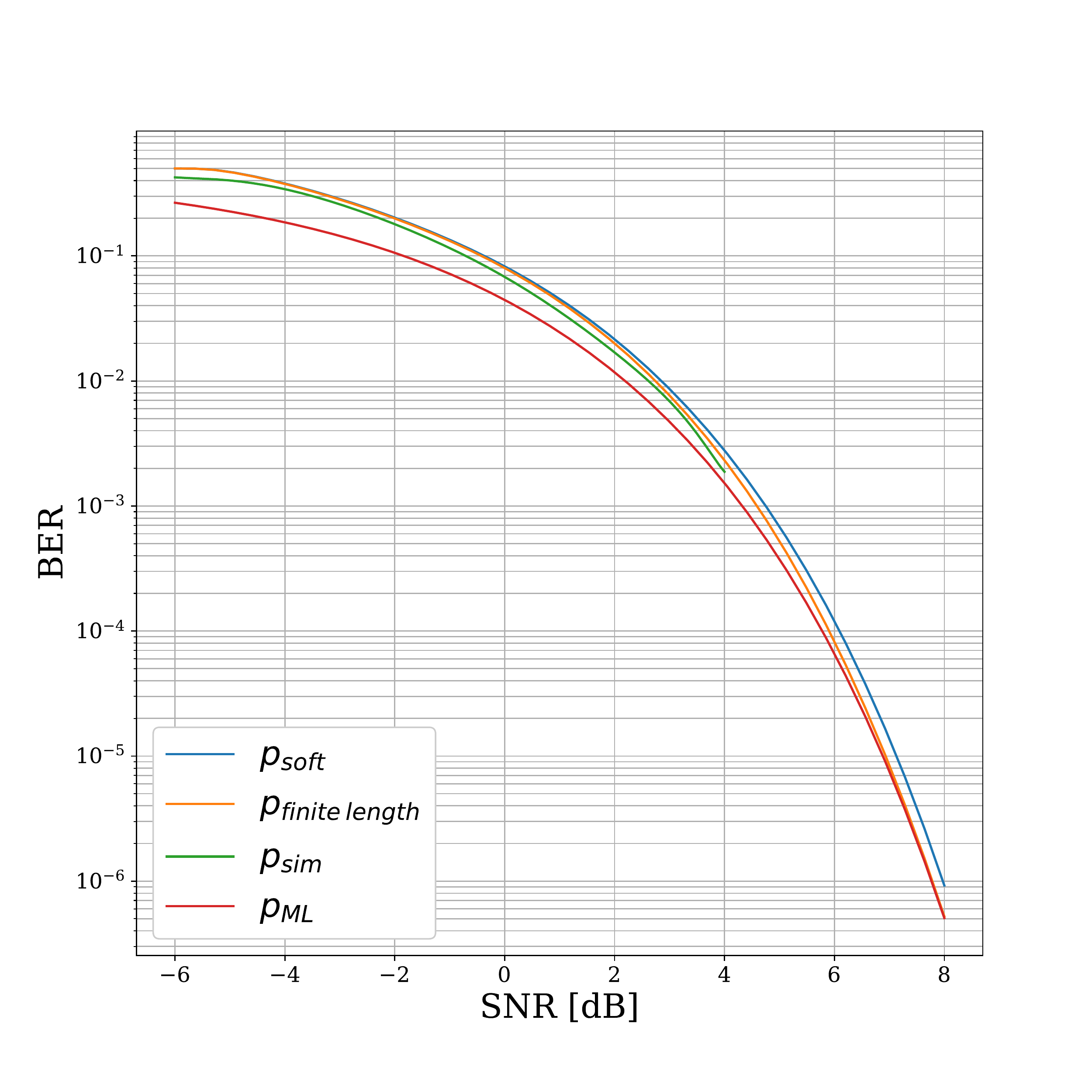}\vspace
{-0.07in} \caption{Simulation results and analytical bounds for the algorithm
$\Psi_{soft}$ applied to modulation-type codes $C_{128}$ of length $8256.$ }%
\label{fig:BER1}%
\end{figure}

In Fig. \ref{fig:BER1} of this section, we will plot analytical bound
(\ref{p-soft}) along with simulation results and the lower bound
(\ref{event-2}) of ML decoding. We will see that all three bounds of Fig.
\ref{fig:BER1} give very tight approximations.

We begin the proof of Theorem \ref{lm:4} with Lemma \ref{lm:1-5}. Here we
analyze the sums of r.v. $z_{j}$ that have asymptotic pdf $\mathbb{N}%
(\delta,\delta)$ with a small bias $\delta\rightarrow0.$

\begin{lemma}
\label{lm:1-5} Consider $m$ independent r.v. $z_{1},...,z_{m}$ with pdf
$\mathbb{N}(\delta,\delta),$ where $\delta\sim c/m.$ Let $Z=\sum
\nolimits_{j}z_{j}$ and $Y=\sum\nolimits_{j}z_{i.j}^{2}.$ Then for
$m\rightarrow\infty,$
\begin{equation}
E\left(  Z\,|\,Y\right)  \sim E\left(  Z\right)  \sim c \label{indep}%
\end{equation}

\end{lemma}

\noindent\textit{Proof. }Consider\textit{ } r.v. \textit{\ }$\varepsilon
_{j}=z_{j}-\delta$ that has pdf $\mathbb{N}(0,\delta).$ Let $R=\sum
_{j}\varepsilon_{j}^{2}.$ This $r.v.$ has $\aleph^{2}$ distribution that tends
to $\mathbb{N}(c,2\delta c)$ as $m\rightarrow\infty.$ Next, note that r.v.
$z_{j}^{2}$ and $\varepsilon_{j}^{2}$ are equivalent with high probability.
Indeed,
\begin{equation}
z_{j}^{2}=\varepsilon_{j}^{2}+2\delta\varepsilon_{j}+\delta^{2}\sim
\varepsilon_{j}^{2} \label{indep-2}%
\end{equation}
Here with high probability we have two events. First, $\varepsilon_{j}^{2}%
\geq\sqrt{\delta}/\ln m,$ whereas the terms $\left\vert \delta\varepsilon
_{j}\right\vert $ and $\delta^{2}$ are bounded from above by $\delta^{3/2}\ln
m=o\left(  \sqrt{\delta}/\ln m\right)  .$ Thus, $z_{j}^{2}\sim\varepsilon
_{j}^{2}$ and $Y\sim R$ as $m\rightarrow\infty.$ In turn, this implies that
r.v. $Y_{i}$ has asymptotic pdf $\mathbb{N}(c,2\delta c)$.

To prove (\ref{indep}), we now may consider unbiased r.v. $\varepsilon_{j}$
and prove asymptotic equality
\begin{equation}
E\left(  \sum\nolimits_{j}\varepsilon_{j}\,|\,R\right)  \sim E\left(
\sum\nolimits_{j}\varepsilon_{j}\right)  =0 \label{indep-1}%
\end{equation}
Consider any subset $\mathcal{S}$ of $2^{m}$ unbiased vectors $(\pm
\varepsilon_{1},...,\pm\varepsilon_{m})$ that give the same sum $R=\sum
_{j}\varepsilon_{j}^{2}.$ Then asymptotic equality (\ref{indep-1}) holds for
each subset $\mathcal{S},$ which proves Lemma \ref{lm:1-5}.\textit{\ }%
\hfill${\dimen0=1.5ex\advance\dimen0by-0.8pt\relax\blacksquare}$\smallskip

To prove Theorem \ref{lm:4}, we will first study r.v. $u_{i\,|\,\ell}$ and
their average \textit{power} moments
\begin{align}
x_{\ell}  &  =E\sum\nolimits_{i}\left(  u_{i\,|\,\ell}/m\right) \label{u-1}\\
\sigma_{\ell}^{2}  &  =E\sum\nolimits_{i}\left(  u_{i\,|\,\ell}^{2}/m\right)
\label{u-2}%
\end{align}
\textit{ } Then r.v. $u_{\ell}=\sum\nolimits_{i}u_{i\,|\,\ell}/m$ has power
moments $x_{\ell}$ and $\sigma_{\ell}^{2}/m$ (here we assume that r.v.
$u_{i\,|\,\ell}$ are weakly dependent).

In the following statements (Lemmas \ref{lm:1-3}-\ref{lm:1-4} and Theorem
\ref{lm:4}), we will show that r.v. $u_{\ell}$ undergo two different processes
as $\ell\rightarrow\infty.$ In the initial iterations $\ell=1,...,$ r.v.
$u_{\ell}$ take vanishing values with high probability as $m\rightarrow
\infty.$ In these iterations, they also may take multiple random walks across
the origin. For $c<1$ and $\ell\rightarrow\infty,$ r.v. $u_{\ell}$ converge to
0. By contrast, for $c>1,$ r.v. $u_{\ell}$ gradually move away from the origin
in opposite directions, albeit with different probabilities. In the process,
r.v. $u_{\ell}$ cross 0 with the rapidly declining probabilities as
$\ell\rightarrow\infty.$ They approach two end points, $x_{\ast}$ and
$-x_{\ast}$ with probabilities $1-P_{\infty}$ and $P_{\infty},$ respectively,
and converge to these points after $\ell\gtrsim\ln m/\ln c$ iterations. At
this point, any r.v. $u_{i\,|\,\ell}$ (that represents a specific bit $i)$ has
BER of $Q\left(  \sqrt{x_{\ast}c}\right)  $ and $1-Q\left(  \sqrt{x_{\ast}%
c}\right)  $. This constitutes bound (\ref{p-soft}).

We first derive how quantities $x_{\ell}$ and $\sigma_{\ell}^{2}$ change in
consecutive iterations. Let $\sigma>0$ and $-\sigma\leq x\leq\sigma.$ Below we
use two functions
\begin{align}
F_{c}(x,\sigma)  &  =\left(  2\pi\right)  ^{-1/2}\int_{-\infty}^{\infty}%
\tanh\left(  \sigma t\sqrt{c}\right)  e^{-(t-x\sqrt{c}/\sigma)^{2}%
/2}dt\label{f-1}\\
G_{c}(x,\sigma)  &  =\left(  2\pi\right)  ^{-1/2}\int_{-\infty}^{\infty}%
\tanh^{2}\left(  \sigma t\sqrt{c}\right)  e^{-(t-x\sqrt{c}/\sigma)^{2}/2}dt
\label{g-1}%
\end{align}

\begin{lemma}
\label{lm:1-3} Let r.v. $u_{i\,|\,\ell},$ $i=1,..,m,$ have average power
moments $x_{\ell}$ and $\sigma_{\ell}^{2}$ of (\ref{u-1}) and (\ref{u-2}).
Then any r.v. $u_{i\,|\,\ell+1}$ has conditional power moments
\begin{align}
E\left(  x_{\ell+1}\,|\,x_{\ell},\sigma_{\ell}\right)   &  =F_{c}\left(
x_{\ell},\sigma_{\ell}\right) \label{f-2}\\
E\left(  \sigma_{\ell+1}^{2}\,|\,x_{\ell},\sigma_{\ell}\right)   &
=G_{c}\left(  x_{\ell},\sigma_{\ell}\right)  \label{g-a}%
\end{align}

\end{lemma}

\noindent\textit{Proof. } Below we consider r.v. $z_{i,j},$ $Z_{i}%
=\sum\nolimits_{j}z_{i,j}$ and $Y_{i}=\sum\nolimits_{j}z_{i,j}^{2}.$ The proof
of Lemma \ref{lm:1-5} shows that these r.v. have pdfs $\mathbb{N}%
\mathcal{(}\delta,\delta),$ $\mathbb{N}(c,c),$ and $\mathbb{N}(c,2\delta c),$
respectively. For $m\rightarrow\infty,$ we will use three restrictions, all of
which hold with high probability. Firstly, $\left\vert z_{i,j}\right\vert
\leq\Delta,$ where $\Delta=2\sqrt{\delta}\ln m\rightarrow0.$ Indeed,
\begin{equation}
\Pr\left\{  \left\vert z_{i,j}\right\vert >\Delta\right\}  \leq2Q(2\ln
m-\sqrt{\delta})=m^{-2\ln m+o(1)} \label{max-1}%
\end{equation}
Also,
\begin{equation}
c-\sqrt{c\ln m}\leq Z_{i}\leq c+\sqrt{c\ln m} \label{max-0}%
\end{equation}%
\begin{equation}
Y_{i}\in(c-\Delta_{1},c+\Delta_{1}),\;\Delta_{1}=m^{-1}c\ln m \label{max}%
\end{equation}
Since $z_{i,j}\rightarrow0$ for all $i,j$, algorithm $\Psi_{soft}$ can use the
following approximations%
\begin{align}
u_{i\,|\,\ell+1}(j)  &  =u_{i,j}u_{j\,|\,\ell}\sim z_{i,j}u_{j\,|\,\ell
}\label{h-3}\\
h_{i\,|\,\ell+1}(j)  &  =2\tanh^{-1}\left[  z_{i,j}u_{j\,|\,\ell}\right]
\sim2z_{i,j}u_{j\,|\,\ell} \label{h-4}%
\end{align}
Here we assume that r.v. $z_{i,j}$ and $u_{j\,|\,\ell}$ are \textquotedblleft
weakly dependent". Indeed, any estimate of $u_{j\,|\,\ell}$ includes $m-1$
terms and only one term includes r.v. $z_{i,j}.$ We then fix the sums
$Z_{i}=\sum\nolimits_{j}z_{i,j}$ and consider conditional r.v. $z_{i,j}%
u_{j\,|\,\ell}\,|\,Z_{i}.$ Given restrictions (\ref{max-0}) and (\ref{max}) we
obtain the moments
\begin{align}
E\left(  z_{i,j}u_{j\,|\,\ell}|\,Z_{i}\right)   &  =E\left(  z_{i,j}\right)
E(u_{j\,|\,\ell})=x_{\ell}Z_{i}/m\label{e-3}\\
\mathcal{D}\left(  z_{i,j}u_{j\,|\,\ell}\,|\,Z_{i}\right)   &  =E(z_{i,j}%
^{2}\,|\,Z_{i})E(u_{j\,|\,\ell}^{2})-\left(  x_{\ell}Z_{i}/m\right)  ^{2}%
\sim\delta\sigma_{\ell}^{2} \label{u-3}%
\end{align}
Similarly to the proof of Lemma \ref{lm:1-5}, we consider r.v. $z_{i,j}^{2}$
and the sums $Z_{i}$ to be independent. We also remove the term $\left(
x_{\ell}Z_{i}/m\right)  ^{2}$ in (\ref{u-3}). Indeed, this term is immaterial
since $x_{\ell}^{2}\leq\sigma_{\ell}^{2}$ and $\left(  Z_{i}/m\right)
^{2}\lesssim cm^{-2}\ln m=o\left(  \delta\right)  ,$ according to
(\ref{max-0}). In essence, here r.v. $z_{i,j}u_{j\,|\,\ell}$ have negligible
means, which yield similar values of conditional variances $\mathcal{D}\left(
z_{i,j}u_{j\,|\,\ell}\,|\,Z_{i}\right)  $ and the second moments $E\left(
z_{i,j}u_{j\,|\,\ell}\,|\,Z_{i}\right)  ^{2}$.

We can now proceed with r.v. $h_{i\,|\,\ell+1}=2\sum\nolimits_{j}%
z_{i,j}u_{j\,|\,\ell}$ $\,$ that sums up independent r.v. $z_{i,j}%
u_{j\,|\,\ell}$ derived in Step $B$ of $\Psi_{soft}.$ Here we obtain
\begin{gather}
E\left(  h_{i\,|\,\ell+1}|\,Z_{i}\right)  =mE\left(  z_{i,j}u_{j\,|\,\ell
}\,|\,Z_{i}\right)  \sim2x_{\ell}Z_{i}\label{h-1}\\
\mathcal{D}\left(  h_{i\,|\,\ell+1}|\,Z_{i}\right)  =m\mathcal{D}\left(
z_{i,j}u_{j\,|\,\ell}\,|\,Z_{i}\right)  \sim4c\sigma_{\ell}^{2} \label{h-2}%
\end{gather}
We can now proceed with the r.v. $u_{i\,|\,\ell+1}\sim\tanh(h_{i\,|\,\ell
+1}/2)$ used in Step $C$ of $\Psi_{soft}.$ For a given $Z_{i},$ r.v.
$h_{i\,|\,\ell+1}$ has Gaussian pdf $\mathbb{N}(2x_{\ell}Z_{i},4c\sigma_{\ell
}^{2}).$ By using the variables $z\equiv x_{\ell}Z_{i}$ and $t=z/\sigma_{\ell
}\sqrt{c}$, we obtain (\ref{f-2}):%
\begin{gather}
E\left(  u_{i\,|\,\ell+1}\right)  \sim\left(  2\pi\sigma_{\ell}^{2}c\right)
^{-1/2}\int_{-\infty}^{\infty}\tanh(z)e^{-\left(  z-x_{\ell}c\right)
^{2}/2c\sigma_{\ell}^{2}}dz\nonumber\\
=\left(  2\pi\right)  ^{-1/2}\int_{-\infty}^{\infty}\tanh(\sigma_{\ell}%
t\sqrt{c})e^{-\left(  t-x_{\ell}\sqrt{c}/\sigma_{\ell}\right)  ^{2}/2}%
dt=F_{c}(x_{\ell},\sigma_{\ell}) \label{F}%
\end{gather}
Similarly, we obtain (\ref{g-a}):
\begin{equation}
E\left(  u_{i\,|\,\ell+1}^{2}\right)  \sim G_{c}(x_{\ell},\sigma_{\ell})
\label{G}%
\end{equation}
\textit{\ }which completes the proof\textit{. \ }\hfill
${\dimen0=1.5ex\advance\dimen0by-0.8pt\relax\blacksquare}$ \smallskip

Recall that the original r.v. $u_{i\,|\,0}$ have equal power moments
$x_{0}=\sigma_{0}^{2}$ of (\ref{eq-0}). The following lemma shows that
nonlinear transformations (\ref{F}) and (\ref{G}) preserve this equality. It
is for this reason that we rescaled the original r.v. $y_{i,j}$ into $z_{i,j}$
to achieve equality (\ref{eq-0}).

Consider function $F_{c}(x,\sigma)$ of (\ref{f-1}) for $\left\vert
x\right\vert =\sigma^{2}.$ For any $c,$ this gives the function%
\begin{equation}
R_{c}(x)=\left(  2\pi\right)  ^{-1/2}\int_{-\infty}^{\infty}\tanh
(t\sqrt{\left\vert x\right\vert c})e^{-\left(  t-\sqrt{\left\vert x\right\vert
c}\right)  ^{2}/2}dt \label{root-5}%
\end{equation}

\begin{lemma}
\label{lm:1-3b} For any two quantities $x,\sigma$ such that $\left\vert
x\right\vert =\sigma^{2}$ and any $c>0,$ functions $F_{c}(x,\sigma)$ and
$G_{c}(x,\sigma)$ satisfy relation
\begin{equation}%
\begin{array}
[c]{ll}%
F_{c}(x,\sigma)=G_{c}(x,\sigma)=R_{c}(x), & \text{if}\;x\geq0\smallskip\\
F_{c}(x,\sigma)=-G_{c}(x,\sigma)=-R_{c}(x), & \text{if}\;x<0
\end{array}
\label{root-4}%
\end{equation}

\end{lemma}

\noindent\textit{Proof. } Let $x=\sigma^{2}$ and $r=t\sqrt{xc}.$ Then
$e^{-\left(  t-\sqrt{xc}\right)  ^{2}/2}=e^{r}e^{-t^{2}/2}e^{-xc/2}.$ Consider
the function
\[
f(r)=e^{r}\left(  \tanh(r)-\tanh^{2}(r)\right)  =\frac{e^{r}-e^{-r}}%
{1+e^{2r}+e^{-2r}}%
\]
Clearly, $f(r)$ is an odd function of $r.$ Then
\[
F_{c}(x,\sigma)-G_{c}(x,\sigma)=\left(  2\pi xc\right)  ^{-1/2}e^{-xc/2}%
\int_{-\infty}^{\infty}f(r)e^{-r^{2}/2xc}dr=0
\]
The case of $x<0$ is similar. Note that $F_{c}(x,\sigma)$ is an odd function
and $G_{c}(x,\sigma)$ is an even function. Then we proceed as above.
$\hfill\quad{\dimen0=1.5ex\advance\dimen0by-0.8pt\relax\blacksquare}$

\begin{lemma}
\label{lm:1-4} For $c\leq1,$ equation (\ref{eq-1}) has a single solution
$x=0.$ For $c>1,$ equation (\ref{eq-1}) has three solutions: $x=0,$ $x_{\ast
}\in\left(  0,1\right)  $ and $-x_{\ast}.$
\end{lemma}

\noindent\textit{Proof. } Let $x>0.$ Integration in (\ref{root-5}) includes
the pdf of $\mathbb{N}(\sqrt{xc},1),$ which gives negligible contribution
beyond an interval $t\in(-x^{-1/4},x^{-1/4}).$ For $x\rightarrow0,$ we can now
limit \ref{root-5}) to this interval. In this case, $t\sqrt{xc}\rightarrow0$
for any $c$ and $\tanh(t\sqrt{xc})\sim t\sqrt{xc}.$ Then
\begin{equation}
R_{c}(x)\sim\left(  2\pi\right)  ^{-1/2}\int_{-\infty}^{\infty}t\sqrt
{xc}e^{-\left(  t-\sqrt{xc}\right)  ^{2}/2}dt=xc \label{root-6}%
\end{equation}
Thus, inequality $R_{c}(x)>x$ holds for sufficiently small $x$ iff $c>1.$ On
the other hand, $\tanh(t\sqrt{xc})<1$ and therefore $R_{c}(x)<1$ for any $x.$
Now we see that functions $y=R_{c}(x)$ and $y=x$ intersect at some point
$x_{\ast}\in\left(  0,1\right)  $ for any $c>1.$ Finally, it can be verified
that $R_{c}(x)$ has a declining positive derivative $R_{c}^{\prime}(x),$
unlike the constant derivative $1$ of the function $y=x.$ Therefore, equation
(\ref{eq-1}) has a single positive solution $x_{\ast}$. \textit{\ }%
\hfill${\dimen0=1.5ex\advance\dimen0by-0.8pt\relax\blacksquare}$ \smallskip

\begin{figure}[ptbh]
\centering\includegraphics[scale=0.32]{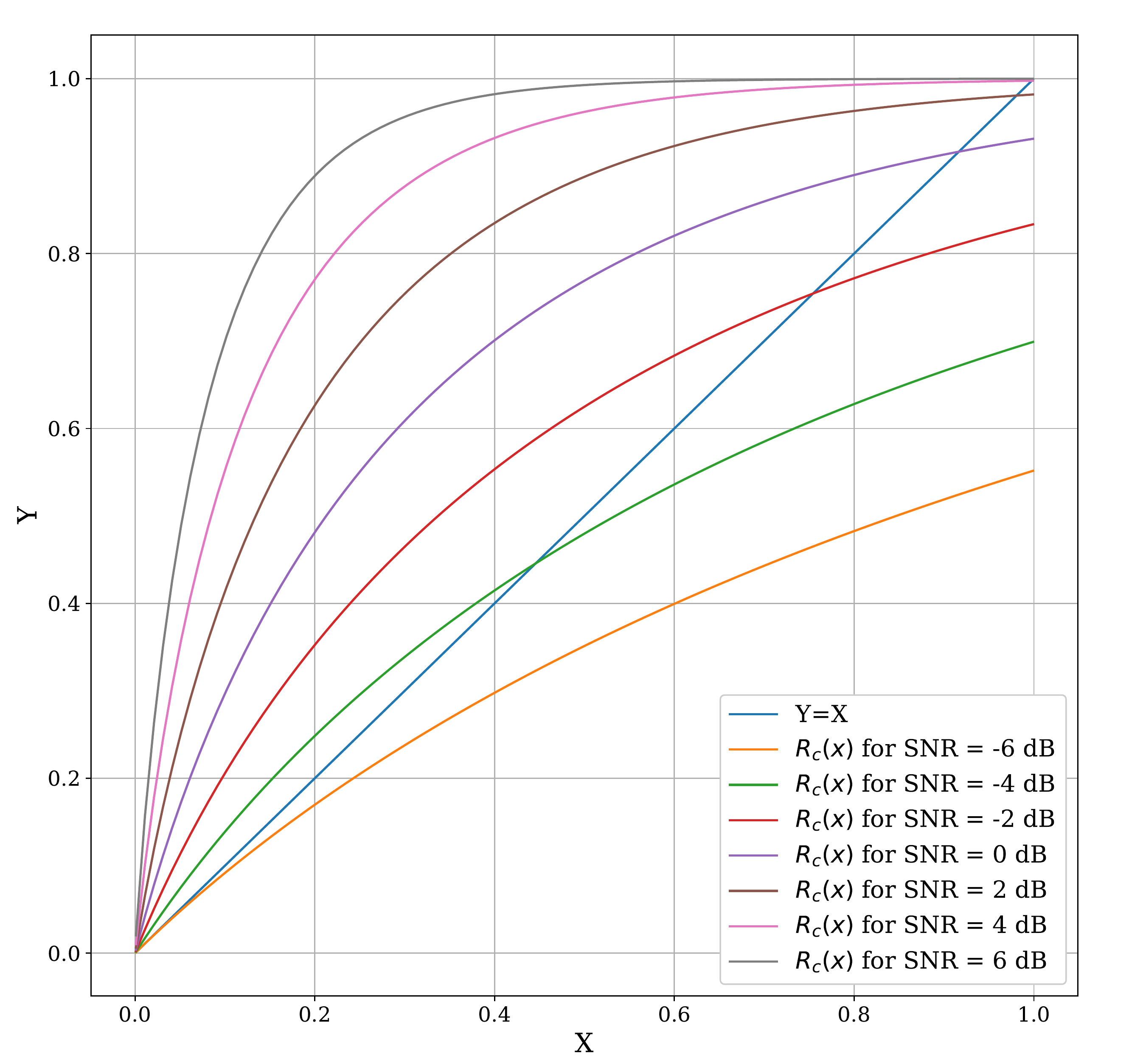}\vspace
{-0.07in} \caption{Functions $y=R_{c}(x)$ and $y=x$ for different values of
$SNR=$ $10\log_{10}\left(  c/4\right)  .$}%
\label{fig:3}%
\end{figure}

In Fig. \ref{fig:3}, function $y=R_{c}(x)$ is shown for different values of
$x\in\lbrack0,1]$ and $SNR=10\log_{10}(c/4).$ The cross-point of functions
$y=R_{c}(x)$ and $y=x$ represents the root $x_{\ast}.$ Here the threshold
$c=1$ corresponds to $SNR=-6$ dB.\vspace{-0.01in}

Summarizing Lemmas \ref{lm:1-3}-\ref{lm:1-4}, we have

\begin{corollary}
\label{cor1} Let $m\rightarrow\infty.$ Then r.v. $u_{i\,|\,\ell},$ $i=1,..,m,$
have power moments $x_{\ell}$ and $\sigma_{\ell}^{2}$ that satisfy equality
$\left\vert x_{\ell}\right\vert =\sigma_{\ell}^{2}$ for any iteration $\ell.$
Iteration $\ell$ transforms $x_{\ell}$ and $\sigma_{\ell}^{2}$ into
\begin{equation}
\left\vert x_{\ell+1}\right\vert =\sigma_{\ell+1}^{2}=R_{c}(x_{\ell})
\label{g-2}%
\end{equation}

\end{corollary}

\textit{Proof of Theorem }\ref{lm:4}.

1. Lemma \ref{lm:1-4} shows that for $c>1,$ function $R_{c}(x_{\ell})$ grows
for positive $x_{\ell}.$ Thus, equality $R_{c}(x_{\ell})=x_{\ell}$ holds iff
$x_{\ell}=x_{\ast}$, where $x_{\ast}$ the root of (\ref{eq-1}). Next, consider
initial iterations $\ell=0,...$ Here r.v. $u_{0}$ has pdf $\mathbb{N}%
(\delta,\delta/m)$ and (with high probability) has vanishing values
$\left\vert u_{0}\right\vert \leq\sqrt{\delta/m}\ln m.$ In further iterations
$\ell,$ transform (\ref{root-6}) performs simple scaling $x_{\ell+1}\sim
cx_{\ell}$ as long as $x_{\ell}\rightarrow0$ for $m\rightarrow\infty.$ Thus,
algorithm $\Psi_{soft}$ fails for $c<1$ since $x_{\ell}\rightarrow0$ in this case.

2. Now let $c>1$ and $L=\ln m\,/\,\ln c.$ Note that $u_{0}<0$ with probability
$Q(\sqrt{\delta m})\sim Q(\sqrt{c}).$ For iterations $\ell=o(L)$ and
$m\rightarrow\infty,$ we still obtain vanishing moments $\left\vert E(u_{\ell
})\right\vert \lesssim c^{\ell}\delta\rightarrow0$ $.$It can also be verified
that $E(u_{\ell})$ moves away from 0 in $\mu=\alpha L$ iterations for some
$\alpha>0$.$.$ Note also that r.v. $u_{\ell}$ has variance $\mathcal{D}%
(u_{\ell})\leq\mathcal{D}(u_{i\,|\,\ell})/m\leq1/m.$ Thus, both cases,
$u_{\ell}\rightarrow x_{\ast}$ or $u_{\ell}\rightarrow-x_{\ast},$ hold with
high probability as $\ell\rightarrow\infty$.

3. We can now derive the BER for both cases. From (\ref{h-1}) and (\ref{u-3}),
we see that the Gaussian random variable $h_{i\,|\,\ell+1}$ has the moments
\[
E\left(  h_{i\,|\,\ell+1}\right)  \sim2x_{\ell}E\left(  Z_{i}\right)
=2x_{\ell}c,\quad\mathcal{D}\left(  h_{i\,|\,\ell+1}\right)  \sim
4c\sigma_{\ell}^{2}%
\]
For any iteration $\ell,$ we can now estimate BER $p_{i\,|\,\ell+1}%
=\Pr\{h_{i\,|\,\ell+1}<0\}$ as
\begin{equation}
p_{i\,|\,\ell+1}=Q\left(  x_{\ell}c/\sigma_{\ell}\sqrt{c}\right)  =\left\{
\begin{array}
[c]{ll}%
Q\left(  \sqrt{x_{\ell}c}\right)  , & \text{if}\;x_{\ell}>0\smallskip\\
1-Q\left(  \sqrt{-x_{\ell}c}\right)  , & \text{if}\;x_{\ell}<0
\end{array}
\right.  \label{pr}%
\end{equation}

4. Consider the probabilities $P_{\ell}=\Pr\left\{  x_{\ell}<0\right\}  $ and
$1-P_{\ell}=\Pr\left\{  x_{\ell}>0\right\}  ,$ which define conditions of
(\ref{pr}). We will now use two partial distributions of r.v. $u_{\ell}$ that
have opposite means $\pm b_{\ell},$ where $b_{\ell}=|x_{\ell}|.$ According to
(\ref{g-2}), r.v. $u_{i\,|\,\ell}$ have the second moment $E(u_{i\,|\,\ell
}^{2})=b_{\ell}.$ Then r.v. $u_{\ell}=\sum\nolimits_{i}\left(  u_{i\,|\,\ell
}/m\right)  $ has the pdf $\mathbb{N}\mathcal{(}\pm b_{\ell},\eta_{\ell})$
with the variance
\[
\eta_{\ell}=\left(  b_{\ell}-x_{\ell}^{2}\right)  /m=b_{\ell}(1-b_{\ell})/m
\]
Note that $b_{\ell}\rightarrow x_{\ast}$ for $\ell>L,$ whereas $\eta_{\ell
}\rightarrow0$ as $\ell,m\rightarrow\infty.$ Thus, r.v. $u_{\ell}$ cross $0$
with a vanishing probability for any iteration $\ell>L.$ On the other hand,
r.v. $u_{\ell}$ may cross 0 multiple times if $\ell=o(L).$ From now on, we
take $\ell=o(L).$ Then we will express $P_{\ell+1}$ via $P_{\ell}$ using the
mean
\[
b_{\ell}=c^{\ell}\delta
\]

5. Consider both distributions $\mathbb{N}\mathcal{(}x_{\ell},\eta_{\ell}),$
where $x_{\ell}=\pm b_{\ell}=\pm c^{\ell}\delta.$ Given some value $u$ of r.v.
$u_{\ell},$ define r.v. $u_{\ell+1}\,|\,u=m^{-1}\sum\nolimits_{i}\left(
u_{i\,|\,\ell+1}\,|\,u\right)  .$ This r.v. has pdf%
\[
p(u)=\mathbb{N}(cu,c\eta_{\ell})=(2\pi\eta_{\ell})^{-1/2}e^{-(u-x_{\ell}%
)^{2}m/2\eta_{\ell}}%
\]
First, let $E(u_{\ell})=b_{\ell}.$ Clearly $\Pr\{cu<0\}=Q(u\sqrt{c/\eta_{\ell
}}).$ Then we average over all values $u$ of $u_{\ell}$ and obtain the
probability
\begin{align*}
S_{\ell}  &  =\Pr\{u_{\ell+1}\,<0\,|\,E(u_{\ell})=b_{\ell}\}=\int_{-\infty
}^{\infty}Q(u\sqrt{c/\eta_{\ell}})p(u)du\\
&  \sim(2\pi)^{-1/2}\int_{-\infty}^{\infty}Q(t\sqrt{c})e^{-(t-b_{\ell}%
/\sqrt{\eta_{\ell}})^{2}/2}dt
\end{align*}
Here we use variable $t=u/\sqrt{\eta_{\ell}}.$ Next, we consider the initial
iterations $\ell=o(\ln m/\ln c)$ and introduce parameter
\begin{equation}
C_{\ell}=b_{\ell}/\sqrt{\eta_{\ell}}\sim\sqrt{c^{\ell+1}/(1-m^{-1}c^{\ell+1}%
)}\sim c^{\left(  \ell+1\right)  /2} \label{c-ell}%
\end{equation}
Note that $b_{\ell}/\sqrt{\eta_{\ell}}=C_{\ell}\sim c_{\ell},$ which gives
(\ref{q-ell}). Similarly, for $E(u_{\ell})=-b_{\ell},$ we obtain the
probability
\[
Q_{\ell}=\Pr\{u_{\ell+1}<0\,|\,E(u_{\ell})=-b_{\ell}\}=\int_{-\infty}^{\infty
}Q(u/\sqrt{c/\eta_{\ell}})p(-u)du
\]
For $\ell<L=\ln m/\ln c,$ this gives the probability
\begin{equation}
P_{\ell+1}=\Pr\left\{  u_{\ell+1}<0\right\}  =\left(  1-P_{\ell}\right)
S_{\ell}+P_{\ell}Q_{\ell}=S_{\ell}+P_{\ell}T_{\ell} \label{rec-a}%
\end{equation}
where $T_{\ell}=Q_{\ell}-S_{\ell}$ is given by (\ref{p-0}). We can also
slightly tighten estimates (\ref{q-ell}) and (\ref{p-0}), by using quantity
$C_{\ell}$ of (\ref{c-ell}) instead of $c_{\ell}.$

We can now proceed with iterations $P_{\ell}$, which begin with $P_{0}%
=Q(\sqrt{c}).$ For any $\ell,$ quantities $S_{\ell}$ and $T_{\ell}$ depend on
$c$ only. Also, quantities $c_{\ell}=c^{\left(  \ell+1\right)  /2}$ grow
exponentially, in which case $S_{\ell}\rightarrow0$ and $Q_{\ell}%
\rightarrow1.$ Thus, quantities $P_{\ell}$ converge, since $P_{\ell+1}\sim
P_{\ell}Q_{\ell}$ for sufficiently large $\ell\geq L.$

We can now evaluate $P_{soft}.$ For $\ell\rightarrow\infty,$ we replace
$P_{\ell}$ with $P_{\infty}$ in (\ref{rec-a}) and use $x_{\ast}$ of
(\ref{eq-1})$.$ Finally, note that (\ref{p-soft}) is only an asymptotic
estimate. Here we excluded the residual term $O\left(  \ln m/\sqrt{m}\right)
$ used in approximations (\ref{max-1}) and (\ref{max}). \hfill
${\dimen0=1.5ex\advance\dimen0by-0.8pt\relax\blacksquare}$

\textit{High-signal case. }Consider functions $S_{\ell}$ and $T_{\ell}$ of
(\ref{q-ell}) and (\ref{p-0}) as $c\rightarrow\infty.$ Then $S_{\ell
}\rightarrow0,$ $T_{\ell}\rightarrow1,$ and $P_{\infty}\rightarrow
P_{0}=Q\left(  \sqrt{c}\right)  .$ In this case, $P_{soft}\sim2Q\left(
\sqrt{c}\right)  \sim(2/\pi c)^{1/2}e^{-c/2}.\ $The latter represents a 3 dB
gain over the uncoded modulation, whose BER has the order of $e^{-c/4}.$
\textit{\ }

\textit{Complexity}. Given $m$ information bits, algorithm $\Psi_{soft}$ has
complexity of order $m^{2}\log m$. Indeed, each iteration $\ell$ recalculates
quantities $u_{i\,|\,\ell}(j)$ and $h_{i\,|\,\ell}(j)$ for all ordered pairs
$(i,j).$ This requires $O(m^{2})$ operations. We also need $O(\log m\,/\,\log
c)$ iterations $\ell$ to make the estimates $u_{i\,|\,\ell}$ bounded away from
0 as $m\rightarrow\infty$. Also, it can be shown that the stable point
$x_{\ast}$ can be reached within a margin $\varepsilon\rightarrow0$ in
$O\left(  \ln\varepsilon^{-1}\,/\,\ln c\right)  \,$ iterations$.$ For
$\varepsilon=m^{-1},$ this gives the overall complexity of $m^{2}\ln m\,/\,\ln
c$ operations.

\textit{Simulation results vs analytical bounds.} In Fig. \ref{fig:BER1}, we
plot analytical bound $P_{soft}$ of (\ref{p-soft}) along with simulation
results $P_{sim}$ and the lower bound $P_{ML}$ of (\ref{event-2}). Here we
consider codes $C_{m}$ of dimension $m=128$ on the AWGN channels with various
SNRs $10\log_{10}(c/4).$ We see that both bounds (\ref{p-soft}) and
(\ref{event-2}) tightly follow simulation results and each other. This also
supports our main assumption that the algorithm $\Psi_{soft}$ can be
considered using independent random variables. For completeness, we also plot
non-asymptotic bound $P_{finite\;length}$ obtained by using parameters
$C_{\ell}$ of (\ref{c-ell}) in both formulas (\ref{q-ell}) and (\ref{p-0}).
Unexpectedly, this bound completely coincides with a much simpler lower bound
$P_{ML}$ for high SNR.

\section{Multilevel protection schemes}

Let $B_{i}=B_{i}(\mu,\mu r_{i}$) be a sequence of $b$ capacity-achieving polar
codes. The rates $0\leq r_{0}<...<r_{b-1}$ will be specified later. We first
encode data block $\overline{\mathbf{a}_{i}}$ of length $\mu r_{i}$ into some
vector $A_{i}\in B_{i}$ and then form a compound block $A=\left(
A_{0},...,A_{b-1}\right)  $ of length $m=\mu b.$ Below $\mu\rightarrow\infty$
and $\ b$ is a constant. Block $A$ is further encoded by code $C_{m}$ of rate
$R_{m}=2/(m+1)$ and length $n=\left(  _{2}^{m+1}\right)  .$ We use notation
$\widehat{C}_{m}$ for the compound code of rate $R\sim R_{m}r,$ where
$r=\sum\nolimits_{i}r_{i}/b.$ Thus, code $\widehat{C}_{m}$ reduces code rate
$R_{m}$ by a factor of $r,$ which gives SNR of $c/4r$ per information bit$.$

Let $I_{s}=\{\mu s+1,...,\mu(s+1)\}$ for any $s=0,...,b-1.$ The received block
$\widehat{C}=\widehat{C}(0)$ of length $n$ is first decoded by the algorithm
$\Psi_{soft}$ using $L=O(\ln m)$ iterations$.$ The result is some block
$\widehat{A}(0)$ of length $m.$ We then retrieve the first $\mu$ decoded bits
in $\widehat{A}(0)$ that form the sub-block $\widehat{A}_{0}=\left(
\widehat{a}_{1},...,\widehat{a}_{\mu}\right)  $ of length $\mu.$ Block
$\widehat{A}_{0}$ is decoded by a polar code $B_{0}$ into some block
$A_{0}=\{a_{1},...,$ $a_{\mu}\}.$ We assume that the corrected block $A_{0}$
has $WER\rightarrow0$ as $\mu\rightarrow\infty$. We then use $A_{0}$ to
replace the first $\mu$ symbols of the block $\widehat{C}(0)$. The result is a
new block $\widehat{C}(1)$ of length $n.$ This completes round $s=0$.

Round $s=1$ is similar. Algorithm $\Psi_{soft}$ now also employs block $A_{0}$
to recalculate the remaining $m-\mu$ information bits of $\widehat{C}(1).$ The
obtained sub-block $\widehat{A}_{1}=\left(  \widehat{a}_{\mu+1},...,\widehat
{a}_{2\mu}\right)  $ is decoded into some vector $A_{1}=\{a_{\mu+1},...,$
$a_{2\mu}\}$ using code $B_{1}.$ Then $A_{1}$ replaces $\widehat{A}_{1}$ in
positions $i\in I_{1}$ and yields a new block $\widehat{C}(2)$. Similarly,
rounds $s=2,...,b-1$ only retrieve a block $A_{s}$ on positions $i\in I_{s}$
Then we obtain block $\widehat{C}(s+1)$ that include corrected bits
$a_{1},...,$ $a_{\left(  s+1\right)  \mu}.$

In any round $s,$ $\mu s$ corrected information bits serve as frozen bits and
aid the algorithm $\Psi_{soft}$. Indeed, with high probability, we use correct
estimates $u_{j\,|\,\ell}=a_{j}$ for all $j\leq\mu s.$ Then the parity checks
$u_{i\,|\,\ell+1}(j)=u_{i,j}u_{j\,|\,\ell}$ are reduced to the
repetitions/inversions $u_{i\,|\,\ell+1}(j)=a_{j}u_{i,j}$ of symbols
$u_{i,j}.$ Also, recall that algorithm (\ref{soft}) outputs the likelihoods
$h_{i\,|\,L}$ of all symbols $a_{i}.$ Thus, we use $h_{i\,|\,L}$ as our bit
estimates in every round $s$ as follows.
\[
\frame{$%
\begin{array}
[c]{l}%
\text{For all }i\in\{\mu s+1,...,m\}\text{ and }j\in\{1,...,m\}:\smallskip\\
A\text{. Use block }\widehat{C}(s).\text{ Derive }u_{i\,|\,\ell+1}%
(j)=u_{i,j}u_{j\,|\,\ell}\smallskip\\
\quad\text{and }h_{i\,|\,\ell+1}(j)=2\tanh^{-1}(u_{i,j}u_{j\,|\,\ell
})\smallskip\\
B\text{. Derive }h_{i\,|\,\ell+1}=\sum\nolimits_{j}h_{i\,|\,\ell+1}(j)\text{
}\smallskip\\
C\text{. If }\ell<L,\text{ find }u_{i\,|\,\ell+1}=\tanh\left(  h_{i\,|\,\ell
+1}/2\right)  .\smallskip\\
\quad\text{Goto A with }u_{i\,|\,\ell+1}\text{ and }\ell:=\ell+1.\smallskip\\
D\text{. If }\ell=L,\text{ use block }\widehat{A}_{s}=(h_{i\,|\,L},\text{
}i\in I_{s}).\smallskip\\
\quad\text{Decode it into }A_{s}\in B_{s}(\mu,\mu r_{s}).\smallskip\\
E.\text{ Replace }\widehat{A}_{s}\text{ with }A_{s}\text{ to form }\widehat
{C}(s+1).\smallskip\\
\text{If }s<b-1,\text{let }s:=s+1,\text{ }\ell:=0.\text{ Goto }A.\smallskip\\
\text{If }s=b-1,\text{ output bits }a_{1},...,a_{m}.\smallskip
\end{array}
$ }%
\]
Let an information block $A$ consist of $m$ zeros. We then use antipodal
signaling and transmit a codeword $1^{n}$ over an AWGN channel$.$ Round $s$
includes $\mu s$ correct information bits $u_{i\,|\,\ell}=a_{i}=1.$ Let
$\lambda_{s}=s/b.$ Then the remaining $m-\mu s$ r.v. $u_{i\,|\,\ell}$, $i>\mu
s,$ have the average power moments\textit{\ }
\begin{align}
x_{\ell}  &  =\left[  m\left(  1-\lambda_{s}\right)  \right]  ^{-1}%
\sum\nolimits_{i>\mu s}Eu_{i\,|\,\ell}\label{x-1}\\
\sigma_{\ell}^{2}  &  =\left[  m\left(  1-\lambda_{s}\right)  \right]
^{-1}\sum\nolimits_{i>\mu s}E\left(  u_{i\,|\,\ell}^{2}\right)  \label{s-1}%
\end{align}
In particular, the initial setup with $\ell=0$ employs the original r.v.
$u_{i\,|\,0}$ that have asymptotic pdf $\mathbb{N}(\delta,\delta)$ for all
$i>\mu s$ and satisfy equalities $x_{0}=\sigma_{0}^{2}=\delta.$

\begin{theorem}
\label{lm:5} Let the algorithm $\Psi_{soft}$ have $\lambda m$ correct
information symbols $a_{1}=...=a_{\lambda m}=1,$ where $\lambda\in(0,1).$ Then
the remaining $(1-\lambda)m$ symbols $a_{i}$ have BER\
\begin{equation}
P_{soft}(\lambda,c)\sim Q\left(  \sqrt{cX(\lambda)}\right)  \label{eq-3}%
\end{equation}
where $X(\lambda)$ satisfies equations%
\begin{align}
X(\lambda)  &  =\lambda+\left(  1-\lambda\right)  x(\lambda)\label{eq-4a}\\
x(\lambda)  &  =\left(  2\pi\right)  ^{-1/2}\int_{-\infty}^{\infty}%
\tanh\left(  t\sqrt{cX(\lambda)}\right)  e^{-\left(  t-\sqrt{cX(\lambda
)}\right)  ^{2}/2}dt \label{eq-4}%
\end{align}

\end{theorem}

\noindent\textit{Proof. } In essence, we follow the proof of Theorem
\ref{lm:4}. The main difference - that simplifies the current proof - is that
the former vanishing point $x_{0}=\delta\rightarrow0$ is now replaced with
$X_{0}\rightarrow\lambda.\ $ This removes the random walks across 0 analyzed
in parts 4 and 5 of the former proof. Thus, now we have the case of
$P_{\infty}=0.$ The details are as follows.

For any $j\geq\mu s+1,$ we use approximations (\ref{h-3}) and (\ref{h-4}) and
take $u_{j\,|\,\ell}=1$ for $j\leq\mu s.$ Then
\[
h_{i\,|\,\ell+1}(j)\sim2u_{i\,|\,\ell+1}(j)\sim\left\{
\begin{array}
[c]{ll}%
z_{i,j}u_{j\,|\,\ell}, & \text{if}\;j\geq\mu s+1\smallskip\\
z_{i,j}, & \text{if}\;j\leq\mu s
\end{array}
\right.
\]
For any given $Z_{i},$ consider the sums $Z_{i}^{\prime}=\sum\nolimits_{j\leq
\mu s}z_{i,j}$ and $Z_{i}^{\prime\prime}=\sum\nolimits_{j\geq\mu s+1}z_{i,j}.$
These sums have expected values $E(Z_{i}^{\prime})=\lambda Z_{i}$ and
$E(Z_{i}^{\prime\prime})=\left(  1-\lambda\right)  Z_{i}.\ $Let%
\begin{align*}
X_{\ell}  &  =\lambda+\left(  1-\lambda\right)  x_{\ell}\\
\theta_{\ell}^{2}  &  =\lambda+\left(  1-\lambda\right)  \sigma_{\ell}^{2}%
\end{align*}
Then we define the moments
\begin{gather}
E\left(  h_{i\,|\,\ell+1}\right)  \sim2x_{\ell}Z_{i}^{\prime\prime}%
+2Z_{i}^{\prime}\sim2Z_{i}\left[  \lambda+x_{\ell}\left(  1-\lambda\right)
\right]  =2Z_{i}X_{\ell}\label{h-1a}\\
\mathcal{D}\left(  h_{i\,|\,\ell+1}\right)  \sim4c\left(  1-\lambda\right)
\sigma_{\ell}^{2}+4c\lambda=4c\theta_{\ell}^{2} \label{h-2a}%
\end{gather}
Thus, r.v. $h_{i\,|\,\ell+1}/2$ has Gaussian pdf $\mathbb{N}(X_{\ell}%
c,\theta_{\ell}^{2}c).$

Next. consider r.v. $u_{i\,|\,\ell+1}\sim\tanh(h_{i\,|\,\ell+1}/2).$ Similarly
to equalities (\ref{f-2}) and (\ref{g-a}), we have%
\begin{align}
E\left(  u_{i\,|\,\ell+1}\right)   &  \sim\left(  2\pi\theta_{\ell}%
^{2}c\right)  ^{-1/2}\int_{-\infty}^{\infty}\tanh(z)e^{-\left(  z-X_{\ell
}c\right)  ^{2}/2c\theta_{\ell}^{2}}dz=F_{c}(X_{\ell},\theta_{\ell
})\label{h-3a}\\
E[u_{i\,|\,\ell+1}^{2}]  &  \sim\left(  2\pi\theta_{\ell}^{2}c\right)
^{-1/2}\int_{-\infty}^{\infty}\tanh^{2}(z)e^{-\left(  z-X_{\ell}c\right)
^{2}/2\theta_{\ell}^{2}c}dz=G_{c}(X_{\ell},\theta_{\ell})\nonumber
\end{align}
Any round $s=\lambda b$ begins with the initial values $X_{0}(\lambda)$ and
$\theta_{0}^{2}(\lambda)$ that satisfy equalities
\begin{equation}
X_{0}(\lambda)=\theta_{0}^{2}(\lambda)=\lambda+\delta\left(  1-\lambda\right)
\sim\lambda\label{init}%
\end{equation}
which are similar to the former equality $x_{0}=\sigma_{0}^{2}$. Thus, we may
follow the proof of Theorem \ref{lm:4} and obtain equality $F_{c}(X_{\ell
},\theta_{\ell})=G_{c}(X_{\ell},\theta_{\ell})$ for any iteration $\ell.$ Now
we see that $x_{\ell+1}=\sigma_{\ell+1}^{2}$ and $X_{\ell}=\theta_{\ell}^{2}$.
Then for any $\lambda$ and $\ell\rightarrow\infty,$ we use variables
$x(\lambda)$ and $X(\lambda)=\lambda+\left(  1-\lambda\right)  x(\lambda).$
Equalities (\ref{x-1}) and (\ref{h-3a}) then give
\[
x(\lambda)=E\left(  u_{i}^{\infty}\right)  =F_{c}(X(\lambda),\sqrt{X(\lambda
)})
\]
which can be rewritten as (\ref{eq-4}).

This also gives estimate (\ref{eq-3}). Indeed, iterations (\ref{h-1a}) and
(\ref{h-2a}) show that the original iteration for $\ell=0$ gives r.v.
$h_{i}^{1}$ that has Gaussian pdf $\mathbb{N}(2\lambda c,4\lambda c).$ Then
for any round $s=\lambda b,$ r.v. $u_{1}=m^{-1}\sum\nolimits_{i>\mu s}$
$u_{i\,|\,1}$ has the mean $F_{c}(\lambda c,\lambda c)=R(\lambda c)$ and the
vanishing variance $\mathcal{D=}R(\lambda c)/(1-\lambda)m,$ where $R_{c}(x)$
is defined in (\ref{root-5}). Thus, for any $\lambda>0,$ our iterations begin
with the crossover probability $P_{0}=\Pr\left\{  u_{1}\leq0\right\}
\rightarrow0$ as $m\rightarrow\infty$. The latter implies that $P_{\ell
}\rightarrow0$ for $\ell\rightarrow\infty,$ as defined in (\ref{rec-a}). In
turn, we can remove $P_{\infty}=0$ from (\ref{p-soft}). Now we can use r.v.
$h_{i\,|\,\ell+1}$that have pdf $\mathbb{N}(2X_{\ell}c,4X_{\ell}c),$ according
to (\ref{h-1a}) and (\ref{h-2a}). For $\ell\rightarrow\infty,$ this gives
(\ref{eq-3}) as
\begin{equation}
P_{soft}(\lambda,c)=\Pr\{h_{i\,|\,\infty}<0\}\sim Q\left(  \sqrt{X(\lambda
)c}\right)  \mathit{\ } \label{ch-0}%
\end{equation}

\hfill${\dimen0=1.5ex\advance\dimen0by-0.8pt\relax\blacksquare}$%
\vspace{-0.03in}

\begin{figure}[ptbh]
\centering\includegraphics[scale=0.38]{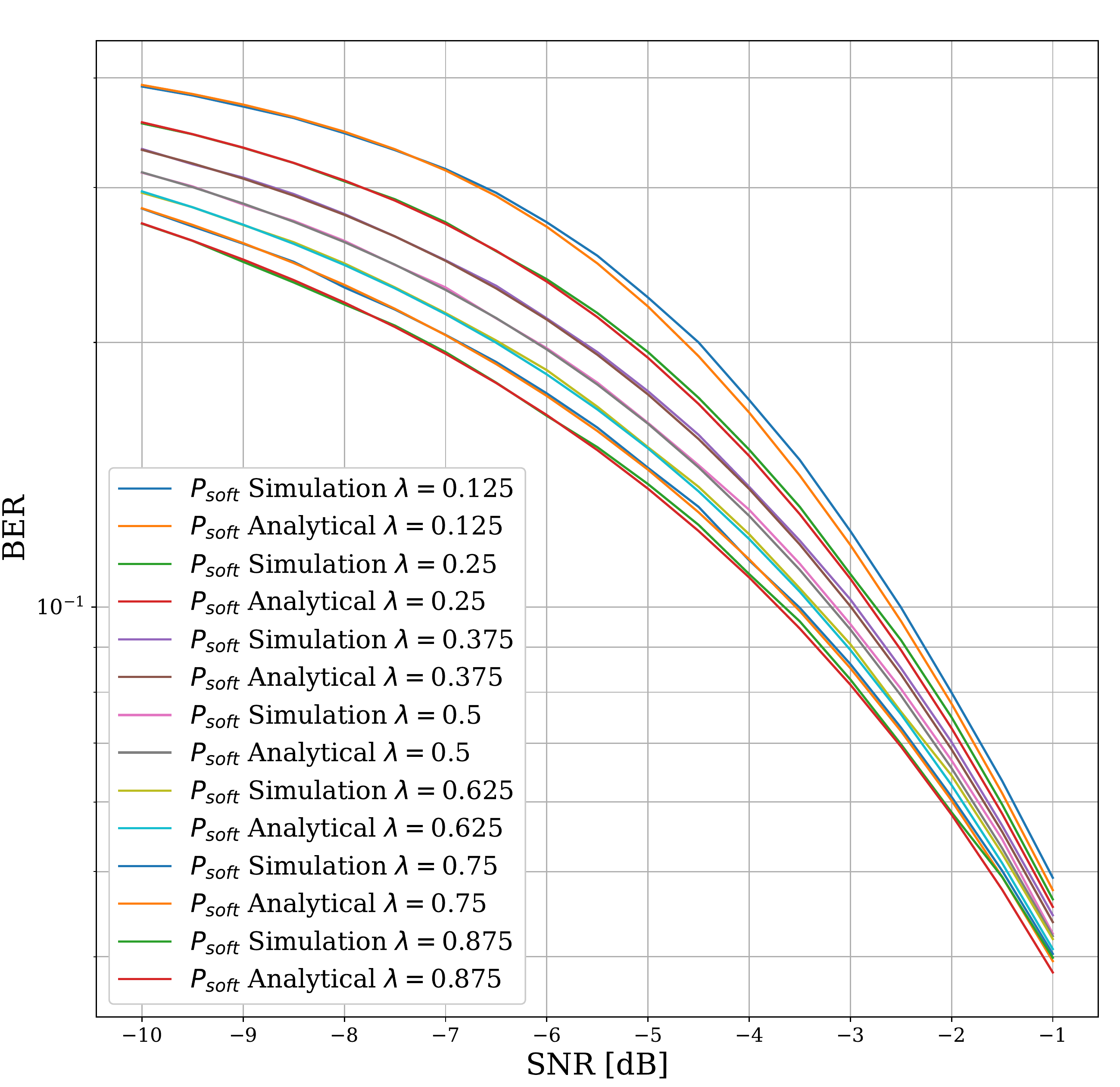}\vspace
{-0.01in} \caption{Simulation results and analytical bounds for the algorithm
$\Psi_{soft}$ applied to modulation-type codes $C_{128}$ with a fraction
$\lambda$ of frozen bits. }%
\label{fig:BER}%
\end{figure}

The absence of random walks in our current setup also makes bound (\ref{eq-3})
very tight. This is shown in Fig. \ref{fig:BER}, where we plot analytical BER
of (\ref{eq-3}) along with simulation results obtained for the algorithm
$\Psi_{soft}(\lambda).$ Here we consider codes $C_{m}$ with $m=128$ and test
various fractions of frozen bits $\lambda=s/m$ and different $S/N$ ratios
$10\log_{10}(c/4).$

\vspace{0in}

Recall that the likelihoods $h_{i\,|\,L}(\lambda)$ give BER (\ref{eq-3}) in
round $s=\lambda b$. We can now represent any Gaussian r.v. $h_{i\,|\,L}%
(\lambda)$ as a channel symbol that has pdf $\mathbb{N(}1,\sigma^{2})$ and a
BER $Q(1/\sigma).$ Thus, $\sigma^{2}=1/cX(\lambda).$ An important note is that
codes $B_{s}(\mu,\mu r_{s})$ now operate on the AWGN channels $\mathbb{N(}%
0,\sigma^{2})$ that have a limited noise power $1/cX(\lambda).$ Unlike the
original code $C_{m},$ we can now use codes $B_{s}(\mu,\mu r_{s})$ of
non-vanishing code rates that grow from $r_{0}$ to $r_{b-1}.$ \vspace{0in}

\begin{theorem}
\label{lm:6} Codes $\widehat{C}_{m}$ of dimension $k\rightarrow\infty$ and
length $n=O(k^{2})$ precoded with $b$ polar codes have overall complexity of
$O(n\ln n).$ For sufficiently large $b,$ these codes achieve a vanishing BER
if used arbitrarily close to the Shannon limit of $-1.5917$ dB per information bit.
\end{theorem}

\noindent\textit{Proof. } In round $s=\lambda b,$ we use a capacity-achieving
code $B_{s}(\mu,\mu r_{s}).$ The corresponding BI-AWGN channel $\mathbb{N}%
_{s}\mathbb{(}0,\sigma_{s}^{2})$ has noise power $\sigma_{s}^{2}=\left(
X(\lambda)c\right)  ^{-1}$ and achieves capacity \cite{AWGN}
\begin{align}
\rho_{c}(\lambda)  &  =\log_{2}\sqrt{\frac{cX(\lambda)}{2\pi e}}-\int
_{-\infty}^{\infty}f(y)\log_{2}f(y)\text{ }dy\label{cap-1}\\
f(y)  &  =\sqrt{\frac{cX(\lambda)}{8\pi}}\left[  e^{-(y+1)^{2}cX(\lambda
)/2}+e^{-(y-1)^{2}cX(\lambda)/2}\right] \nonumber
\end{align}
Here parameter $\lambda$ changes from 0 to 1 in small increments $1/b,$ which
tend to $0$ as $b\rightarrow\infty.$ The average capacity for all AWGN
channels $\mathbb{N}_{s}\mathbb{(}0,\sigma_{s}^{2})$ is $\rho_{c}=\int_{0}%
^{1}\rho_{c}(\lambda)d\lambda.$ Thus, for $m\rightarrow\infty,$ code
$\widehat{C}_{m}$ achieves a vanishing BER for any code rate $r<2\rho_{c}/m,$
which gives $SNR>c/4\rho_{c}.$

We now proceed with code complexity. For $b$ polar codes $B_{s}(\mu,\mu
r_{s}),$ design complexity has the order of $b\mu^{2}\sim2n/b$ or less. Their
decoding requires the order of $b\mu\ln\mu<m\ln m$ operations. Algorithm
$\Psi_{soft}$ includes $b$ rounds with $L=O(\ln m)$ iterations in each round.
This gives complexity order of $n\ln n$ if $b$ is a constant or $n\ln^{2}n$
for growing $b<\ln m.$ Thus, overall complexity has the order of $k^{2}\ln k,$
where $k\rightarrow\rho_{c}m$ is the number of information bits.

To calculate the minimum SNR $\varkappa=\min\nolimits_{c}\left(  c/4\rho
_{c}\right)  ,$ we select parameters $c$ and $b.$ Then we solve equation
(\ref{eq-4a}) for different values of $\lambda=s/b,$ where $s=0,...,b-1$, and
calculate $\rho_{c}$. The following table gives the highest value of code rate
$\rho_{c},$ and the corresponding value of $\varkappa=\varkappa(c,b).$ Here we
count $\varkappa$ in dB, as $10\log_{10}\varkappa$. The last line shows the
gap $\varkappa/\ln2-1$ to the Shannon limit of $\ln2$.\vspace{0in} %

\[%
\begin{tabular}
[c]{|c|c|c|c|c|}\hline
$b$ & $10^{2}$ & $10^{3}$ & $10^{4}$ & $25000$\\\hline
$\rho_{c}$ & $0.404$ & $.3621$ & $.3623$ & $.3623$\\\hline
$\varkappa$ (in dB) & $-1.5655$ & $-1.5890$ & $-1.5915$ & $-1.5917$\\\hline
$\varkappa/\ln2-1$ & $6E-3$ & $7E-4$ & $6E-5$ & $E-5$\\\hline
\end{tabular}
\ \
\]
Finally, note that $b$ is a constant for any $SNR>\ln2.$ Statement \ref{th:1}
now follows directly from the existing bounds \cite{ari} on BER for polar
codes. Here polar codes $B_{i}$ have length $\mu=m/b>2k/b.$ \hfill
${\dimen0=1.5ex\advance\dimen0by-0.8pt\relax\blacksquare}$\vspace{-0.02in}

\section{Concluding remarks}

In this paper, we study new codes that can approach the Shannon limit on the
BI-AWGN channels. We first employ \textquotedblleft modulation " codes $C_{m}$
that use parity checks of weight 3. These codes can be aided by other codes
$B_{m}$ via back-and-forth data recovery. Using BP algorithms that decode
information bits only, codes $C_{m}$ achieve complexity order of $n\ln n$.
Then new analytical techniques give tight lower and upper bounds on the output
BER, which are almost identical to simulation results. Finally, we employ
multilevel codes of dimension $k\rightarrow\infty$ that approach the Shannon
limit with complexity order of $k^{2}$. One open problem is to find out if
there exists a close-form solution to the transcendental equations
(\ref{eq-4a}), which (unexpectedly) give the Shannon limit using numerical
integration in (\ref{cap-1}).

Our future goal is to improve code design for moderate lengths. This work in
progress uses more advanced combinatorial designs for modulation codes. We
conjecture that it also may reduce code complexity to the order of $\ln^{2}k$
operations per information bit for dimensions $k\rightarrow\infty.$

\end{document}